# A Domain Driven Design Based Reference Architecture for IoT Systems Engineering

Yousef Abuseta

Faculty of Information Technology, University of Tripoli, Tripoli, Libya
y.abusetta@uot.edu.ly

**Abstract.** The Internet of Things (IoT) paradigm continues to gain significant traction across diverse do-mains, primarily driven by its transformative and socio-economic benefits. However, engineering robust IoT systems remains a formidable challenge, largely due to the inherent dynamism, volatility, and uncertainty of their operational environments. To mitigate this architectural complexity, microservices have emerged as a viable alternative to traditional, rigid monolithic frameworks, offering a scalable mechanism to fulfill critical quality attributes. To maximize the efficacy of a microservices architecture, its technical boundaries must align precisely with domain-specific business requirements. Domain-Driven Design (DDD) addresses this need by strategically decomposing complex domains into modular, highly manageable subdomains that map directly to individual microservices. Furthermore, to minimize architectural coupling, enhance system integrability, and optimize device energy efficiency, Event-Driven Architecture (EDA) serves as a critical operational enabler. This paper proposes a comprehensive reference architecture for IoT systems engineering, systematically leveraging DDD and EDA principles to harmonize business logic with decoupled, event-driven microservices.

## 1 Introduction

Software systems are primarily engineered to support domain experts—such as medical practitioners, biologists, and financial analysts—in performing complex tasks and alleviating domain-specific challenges. However, a persistent conceptual gap exists: software engineers rarely possess deep knowledge of the target domain, while domain experts are unlikely to be familiar with the nuances of software architecture.

This conceptual gap often results in poorly understood domain requirements, producing misaligned software implementations. Eliciting, modelling, and integrating domain knowledge into the requirements engineering lifecycle remains a formidable challenge, in particular within highly dynamic paradigms such as the Internet of Things (IoT) [1]. The IoT has evolved from a simple system consisting of data-collecting sensors into complex cyber-physical ecosystems, encompassing heterogeneous and distributed components [2]. Engineering software for these environments introduces significant challenges, including addressing heterogeneous hardware protocols, real-time data in-gestion, high-velocity streaming, and evolving business requirements. A significant impediment in traditional IoT software architectures is the issue of the environmental monolith, which represents an architectural anti-pattern where hardware-level device communication is tightly coupled with high-level application business logic. However, empirical research shows that fusing domain-specific context directly into the software engineering lifecycle significantly mitigates requirement ambiguity, improves specification precision, and ensures alignment with complex business processes [3]. Therefore, thoroughly understanding the semantic and operational realities of a target domain is as vital to delivering a successful project as managing the technical aspects of the software development process. Unless structured domain modeling is adopted, requirements engineering often produces brittle specifications that fail to satisfy stakeholder expectations or address operational nuances [4].

To address this socio-technical gap and resolve the linguistic ambiguities that hinder cooperative engineering, Domain-Driven Design (DDD)—originally introduced by Eric Evans [5] for handling enterprise software complexity—has emerged as a foundational methodology in IoT software engineering. The DDD philosophy treats the domain logic as a first-class citizen of the software design process. A cornerstone of this methodology is the Ubiquitous Language concept, which establishes a jargon-free terminology shared universally among technical and non-technical system stakeholders [5]. This common language facilitates the alignment of the underlying software source code, architectural APIs, and database schemas with domain-specific business terminology, thereby alleviating the semantic gap and requirements degradation inherent in complex systems engineering [6].

In addition, DDD introduces Domain Models and Bounded Contexts to address structural complexity. Bounded Contexts explicitly define the conceptual and terminological boundaries within which a specific domain model preserves integrity and maintains structural consistency [5]. By complying with these boundaries, various system components can be developed, maintained, and scaled independently without systemic interference or regression risks. This encapsulation-driven approach allows cross-functional teams to collaborate efficiently, eliminating conflicts across the macro system topology [7].

In modern distributed computing, this structural isolation directly contributes to the success of the microservices architectural style. Increasingly chosen by IoT system designers to achieve crucial quality attributes such as scalability, elasticity, and modifiability, microservices depend largely on the logical boundaries specified by Bounded Contexts [8]. Mapping microservices to bounded contexts yields high functional cohesion, loose coupling, and independent deployment topologies [8]. A key challenge in IoT engineering is defining appropriate microservice boundaries for physical devices and distributed workflows [2]. The alignment of system design directly with physical and operational domains enables engineers to transform monolithic IoT platforms into decoupled, maintainable, and scalable microservices [2].

Despite these theoretical alignments, systematically engineering resilient, microservice-based IoT systems through a synchronized DDD and event-driven approach remains an open research challenge. This paper addresses this gap by formalizing a comprehensive reference architecture for IoT systems engineering utilizing Domain-Driven Design.
The remainder of this paper is organized as follows. Section 2 establishes the necessary background and theoretical foundations of the proposed work. Section 3 contextualizes the study within the current landscape through a review of related literature. Section 4 describes the architectural aspects of the proposed reference architecture in detail. Section 5 concludes the paper and outlines prospective avenues for future research.

# 2 Background

## 2.1 Domain-Driven Design

Domain-Driven Design (DDD) serves as a formal software engineering methodology concentrating on transforming complex business domain models into cohesive, maintainable software architectures [5]. Instead of modelling systems purely around the underlying technical infrastructure such as database schemas or network protocols, DDD treats the business logic as the backbone of the software engineering lifecycle [5]. By setting up collaborative development cycles between technical teams and domain experts, the DDD paradigm systematically fuses domain behaviors, workflow rules, and operational constraints into executable software patterns [6]. The application of DDD is structurally divided into two interdependent phases: Strategic Design and Tactical Design [5, 6].
**Strategic Design**
Strategic design is concerned with the macro-level architectural decomposition process, establishing a techno-structural boundary that correlates system components with high level organizational structures and corporate strategies [5]. It offers the essential blueprint for breaking down complex, monolithic systems into manageable subdomains. The primary building blocks of strategic DDD include:

- **Bounded Contexts**: Precisely defined logical boundaries within which a specific domain model maintains its integrity, remains logically consistent, and operates independently [5].
- **Context Maps**: Graphical structural descriptions that visually establish the integration, data flow, and architectural dependencies between interacting Bounded Contexts [6].
- **Anti-Corruption Layers (ACL)**: Architectural mediators implemented to translate semantics between disparate bounded contexts, forbidding legacy or external domains from contaminating the internal domain model [6].

**Tactical Design**
In contrast, tactical design produces a suite of concrete modeling patterns and programmatic techniques implemented within the boundary of a single Bounded Context [5, 6]. These tactical patterns ensure data integrity, encapsulate business logic, and preserve transactional consistency:

- **Entities**: Uniquely identified domain objects that preserve their identity throughout their lifecycle, regardless of structural changes to their attributes [5].
- **Value Objects**: Immutable elements defined entirely by their structural attributes rather than a distinct identity, frequently used to quantify or describe characteristics of an entity [5].
- **Aggregates**: Cohesive clusters of related Entities and Value Objects tied together by a transactional boundary [6]. External access to the aggregate is routed exclusively through a designated Aggregate Root to enforce internal invariant business rules.
- **Domain Services**: Stateless operations that encapsulate domain logic spanning multiple aggregates. They also represent actions that do not naturally belong within a specific entity or value object [5].
- **Domain Events**: First-lass, immutable artifacts of the domain model that explicitly capture significant state-changing occurrences that require reaction from downstream or adjacent systems [6].

In modern software engineering, this dual-layered approach is increasingly recognized as a vital requirement for distributed system decomposition. In particular, mapping microservice boundaries to strategic Bounded Contexts enforces functional cohesion and loose structural coupling [8, 9], a characteristic highly beneficial when managing the complex, uncertain operational environments of Internet of Things (IoT) frameworks [8].

### 2.2 Event-Driven Architecture (EDA)

Event-Driven Architecture (EDA) is an architectural style designed to enhance performance in asynchronous distributed systems by decoupling components through the production, detection, and consumption of asynchronous state changes, formalized as events [6]. An event represents an immutable historical record of a significant occurrence within a domain boundary [6]. Unlike classical, synchronous request-response mechanisms—which suffer from high temporal coupling and block execution paths—EDA utilizes messaging intermediaries such as message brokers (e.g., MQTT or Apache Kafka) to transmit data asynchronously [6]. This decoupling realizes fundamental quality attributes such as high horizontal scalability, system elasticity, and runtime structural resilience, which are required to steer the fast data flows typical of modern computing environments. [7].

In distributed environments, EDA is employed as a vital structural mechanism for preserving loose coupling across bounded contexts [6]. Within a DDD layout, when a change occurs inside a transactional aggregate boundary, the system avoids invoking adjacent service APIs directly. Instead, the aggregate root raises a Domain Event, publishing it asynchronously to the messaging bus [6]. Downstream bounded contexts intercept this message via an Anti-Corruption Layer (ACL), translating the payload into their local schemas [6]. This integration pattern guarantees that separate distributed services achieve eventual consistency with sub-second latencies without requiring synchronous distributed data locking, effectively preventing system failure cascading across adjacent operational domains [6, 7].

### 2.3 Autonomic Computing and the MAPE-K Feedback Loop

Autonomic computing provides a structured framework for engineering self-adaptive software systems capable of managing themselves with minimal human intervention [10]. To address the operational volatility and unpredictable runtime conditions inherent in modern networks, autonomic architectures externalize adaptation logic from core business operations [10]. This separation of concerns is systematically realized through a continuous closed feedback loop known as the MAPE-K (Monitor, Analyze, Plan, Execute, Knowledge) model, which establishes a structural control relationship between a managing system and a managed system [10].
The autonomic loop is structured into five core functional components that interact sequentially to ensure system stability:

- **Monitor:** Continuously aggregates, filters, and correlates operational telemetry, system logs, and environmental metrics streamed from the managed system infrastructure to maintain an up-to-date observation profile [10].
- **Analyze:** Evaluates the gathered system state indicators against high-level performance objectives, non-functional requirements, and Service Level Agreements (SLAs) to detect behavioral anomalies or service degradation [10].
- **Plan:** Coordinates policy evaluation and reasoning mechanisms to construct an ordered sequence of structural modifications or remediation strategies when anomalies are confirmed [10].
- **Execute:** Implements the calculated change strategies by dispatching precise operational instructions back to physical or virtual effectors within the managed infrastructure without causing service disruption [10].
- **Knowledge:** Serves as the central shared data substrate and memory repository that feeds all four stages of the feedback loop, maintaining runtime architectural models, operational policies, historical logs, and boundary conditions [10, 11].

## 3 Related Work

The industry-wide embrace of the Internet of Things (IoT) has called for an urgent need for software engineering methodologies capable of addressing system complexity and environmental uncertainty. Domain-Driven Design (DDD) has been widely recognized as an efficient approach to handling these structural complexities [5]. Systematic literature reviews on the topic revealed that strategic DDD patterns, such as Bounded Contexts, are highly effective at enhancing software maintainability and supporting developers in defining clean boundaries for microservices [4]. These studies pointed out that successful implementation depends largely on constant collaboration between software developers and domain experts to produce a shared, unambiguous vocabulary known as the Ubiquitous Language [5, 6]. In the context of distributed systems, several researchers have successfully implemented this array of principles. For example, Bogner et al. [8] applied DDD within a microservices architecture to build an IoT monitoring platform capable of hosting multiple distinct projects. Their work showcased that em-ploying DDD to decompose application capabilities yields more precise service definitions, flexible code maintenance, and more adaptable software products [8]. Similarly, Soliman et al. [7] proposed a software framework designed to transform Digital Twins from custom, case-by-case physical model simulations to highly modular microservice structures [7]. By utilizing DDD, their approach formalizes how to handle high data complexity and varied system behaviors without locking the platform into a rigid, non-reusable design [7].
Leveraging these structural foundations, Cabrera et al. [12] proposed MicroIoT, an agile software engineering methodology designed to guide the development of microservice-based IoT applications. By implementing strategic DDD mapping, the MicroIoT framework aligns physical IoT assets and operational processes to distinct Bounded Contexts, producing self-contained microservices that scale independently across distributed networks [12]. To address the underlying device layer, Butzin et al. [13] studied mapping physical sensors and actuators directly to microservices using DDD principles. They pointed out that while traditional enterprise microservices break down monolithic applications top-down, IoT engineering necessitates a bottom-up composition of heterogeneous device services [13]. By modelling device-level microservices around tight, self-contained business capabilities with minimal external dependencies, systems realize higher interoperability and evolvability [13]. Investigating the evolutionary path of these architectures, Shadija et al. [14] examined how microservice architectures differ from traditional Service-Oriented Architecture (SOA) models. Largely influenced by DDD, they determined that microservice boundaries defined by Bounded Contexts enable flexible software platforms capable of scaling effectively to dynamic IoT operational requirements [14]. This structural evolution was evaluated in a practical setting by Wizenty et al. [15], who designed a microservice-based Remote Care Application (RCA) allowing caregivers to interact with smart home components. Utilizing a customized UML profile for DDD, they mapped domain concepts directly into three distinct Bounded Contexts: HistoryContext (for device states and telemetry), RemoteControlContext (for device commands), and AccessControlContext (for user permissions). Their empirical analysis confirmed the direct transformation of DDD domain models into functional microservices, which significantly expedited feature integration and deployment operations.

While standard DDD centers on domain logic and boundary isolation, managing the high-volume data streams inherent in IoT platforms often entails event-driven extensions [6]. Recent studies indicate that combining asynchronous domain events with tactical DDD models significantly diminishes coupling between different parts of a system and improves overall system responsiveness [6]. This combination enables technical infrastructure layers to capture and process telemetry data constantly without neglecting the core business logic of adjacent services [6]. Meanwhile, ensuring that systems remain operational in unpredictable physical environments has resulted in the integration of autonomic feedback loops, such as the MAPE-K (Monitor, Analyze, Plan, Execute, Knowledge) model, into IoT engineering process [10]. Previous work has used decentralized MAPE-K loops to manage runtime adjustments, software migration, and dynamic resource allocation [11]. However, many existing implementations construct these feedback loops using rigid, ad-hoc rules or hard-coded scripts [11]. This often ties the adaptation logic too closely to a specific hardware setup, making it only an application-specific architecture [11].

The current literature revealed that while DDD and autonomic computing have been successfully employed in separate enterprise or simulation problems, their synthesized application within self-adaptive IoT environments remains limited. Most existing studies focus on the implementation details of a single, specific IoT use case, such as a particular data ingestion pipeline or a custom digital twin model. Crucially, there is a clear lack of a formalized, reusable reference architecture that systematically links strategic domain boundaries with autonomic MAPE-K control loops using an event-driven mechanism. Existing approaches do not provide clear guidelines on how to keep transactional device lifecycle operations separated from analytical monitoring data paths. This paper directly addresses this gap by introducing a modular, event-driven reference architecture designed to guide software designers throughout the development lifecycle of resilient, self-adaptive IoT systems.

## 4 Proposed Reference Architecture

This section describes the formalized engineering framework designed to realize self-adaptive, resilient Internet of Things (IoT) ecosystems. Synthesizing the strategic patterns of Domain-Driven Design (DDD) with the operational capabilities of Event-Driven Architecture (EDA), the proposed framework manages system complexity by structurally partitioning an IoT environment into highly cohesive, loosely coupled components.

The architectural lifecycle initiates with the strategic decomposition of the global IoT domain into bounded, manageable subdomains. Each subdomain models a discrete, logically isolated sphere of system expertise. These subdomains are subsequently projected onto distinct Bounded Contexts, within which a strictly defined Ubiquitous Language governs. At the implementation layer, these contexts encapsulate specific tactical DDD building blocks, including Aggregates, Entities, and Value Objects. To maintain systemic structural integrity, control loops, and asynchronous data flows, the operational interactions and structural dependencies between these Bounded Contexts are explicitly formalized via a comprehensive Context Map.

### 4.1 DDD-Based IoT Subdomains and Autonomic Alignment

Modern IoT ecosystems operate within inherently dynamic and unpredictable environments, dictating that they behave as self-adaptive software systems. To enforce the core architectural principle of separation of concerns, the self-adaptation logic must be externalized from the foundational business workflows. Consequently, the proposed reference architecture strategically divides the domain space into two macro-architectural categories: Core IoT Domains (The Managed System) and Self-Adaptation Domains (The Managing System). This structural separation directly mirrors the IBM reference model for autonomic systems, explicitly decoupling the core transactional features of the IoT infrastructure from the feedback loops driving system adaptation.

Information exchange across these domains is driven asynchronously via Domain Events over an EDA messaging substrate to preserve loose coupling and ensure real-time responsiveness. This architecture is depicted in Fig. 1 and detailed as follows.

**Core IoT Domains (Managed System)**

The Core IoT Domains encompass the primary operational layers of the physical and digital infrastructure, managing device topologies and business logic:

- **IoT Device Management Subdomain:** Orchestrates the complete operational lifecycle, registration, remote configuration, and real-time state synchronization of heterogeneous edge devices. It encapsulates physical hardware variations into uniform digital abstractions.
- **Data Ingestion & Processing Subdomain:** Operates as a high-throughput, horizontally scalable ingestion pipeline that processes heterogeneous, high-velocity streaming data emitted by edge networks, ensuring strict data validation and structural normalization.
- **Application-Specific Subdomain:** Encapsulates the core business logic unique to the deployment paradigm—such as Smart Transportation, Smart Agriculture, or Industrial Automation. It represents the primary layer where localized domain rules are programmatically enforced.

**Self-Adaptation Domains (Managing System)**

The Managing System coordinates the runtime modification of the system's topology, behavior, and configurations, mapping directly onto the five pillars of the classical MAPE-K (Monitor, Analyze, Plan, Execute, Knowledge) autonomic loop:

- **Monitoring Subdomain (Monitor):** Continuously aggregates infrastructural, application, and environmental metrics streamlined from the Managed System to establish runtime operational health.
- **Context Management Subdomain (Knowledge):** Acts as the repository for the "Knowledge" element, maintaining global, stateful runtime models, historical trends, operational targets, and environmental contexts.
- **Analysis & Diagnosis Subdomain (Analyze):** Evaluates system performance against QoS objectives, detecting behavioral anomalies or performance bottlenecks and publishing symptom signals when deviations are observed.
- **Adaptation Planning Subdomain (Plan):** Computes optimal, context-aware remediation and adaptation strategies based on received symptom metrics, outputting ordered alteration scripts.
- **Execution & Control Subdomain (Execute):** Translates high-level adaptation plans into concrete control actions, dispatching configurations back to the Core IoT Domains asynchronously to enact live reconfigurations.

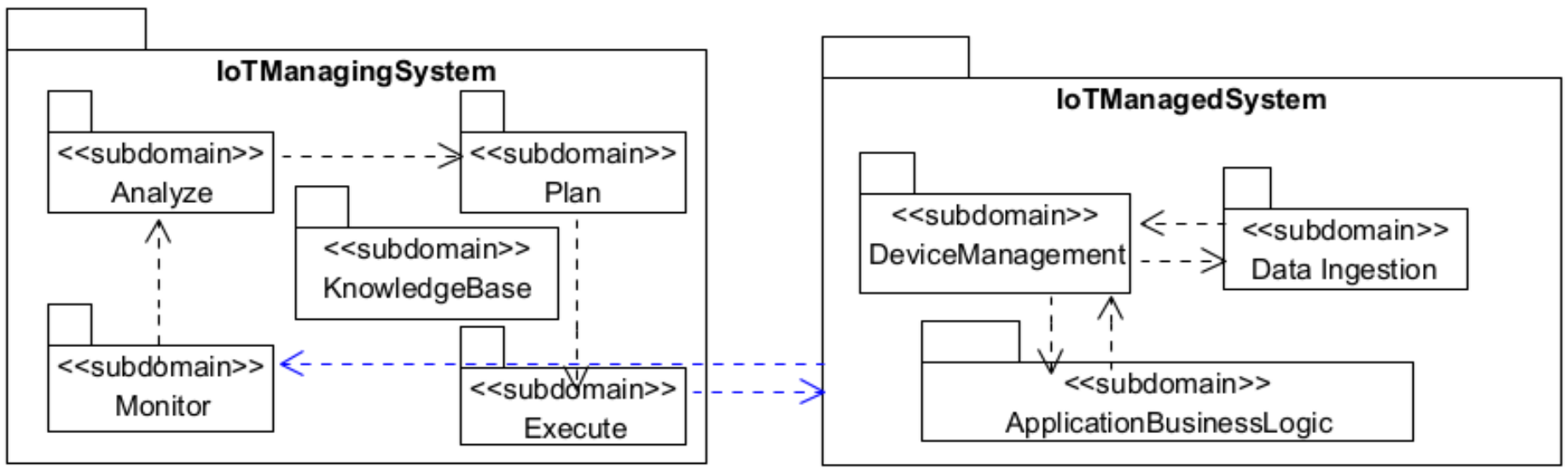

**Fig. 1.** DDD Subdomains for MAPE based IoT Systems.

### 4.2 IoT Bounded Contexts and Tactical Design

To bridge strategic boundaries and software execution, the proposed subdomains are mapped to distinct Bounded Contexts. Each context maintains isolation through bounded transactional models and a localized Ubiquitous Language.

**Device Management Bounded Context**

The Device Management Bounded Context is explicitly concerned with maintaining the structural integrity, health, and operational lifecycle of the physical and virtual device fleet. It encapsulates two primary aggregates to maintain transactional consistency boundaries at scale, namely the Device and Device Group aggregates:

*Device Aggregate*

The Device Aggregate serves as the consistency boundary for an isolated, unique hardware unit.

- **Device Entity (Aggregate Root):** The primary entity representing the physical or virtual device. It holds the unique identifier (DeviceId), tracks current state metrics, and manages core operational access. All external invocations must clear this root to ensure system rules remain intact.
- **Provisioning Entity:** Coordinates device onboarding, cryptographic trust establishment, identity assignment, software installation, and initial state initialization.
- **Digital Twin Entity:** It serves as a control component within the Device aggregate, observing the operational lifecycle, configuration, and active status of the physical nodes. It handles the dual-state synchronization process by detecting the inconsistency between the desired state requested by a system actor (such as an IoT system administrator or the execute component of a MAPE-K control loop) and the reported state transmitted by the physical node. To maintain system stability, the Digital Twin entity validates business invariants immediately within the device aggregate boundary. For instance, a new configuration is rejected unless it complies with the physical constraints specified in the accompanying HardwareSpec Value Object. Whenever a state change occurs, this entity raises an update signal through the aggregate root (the Device entity), activating the publication of a DeviceStateChanged domain event to the underlying messaging bus.

In Fig. 2, A UML package diagram is used to depict the Device Management Bounded Context.

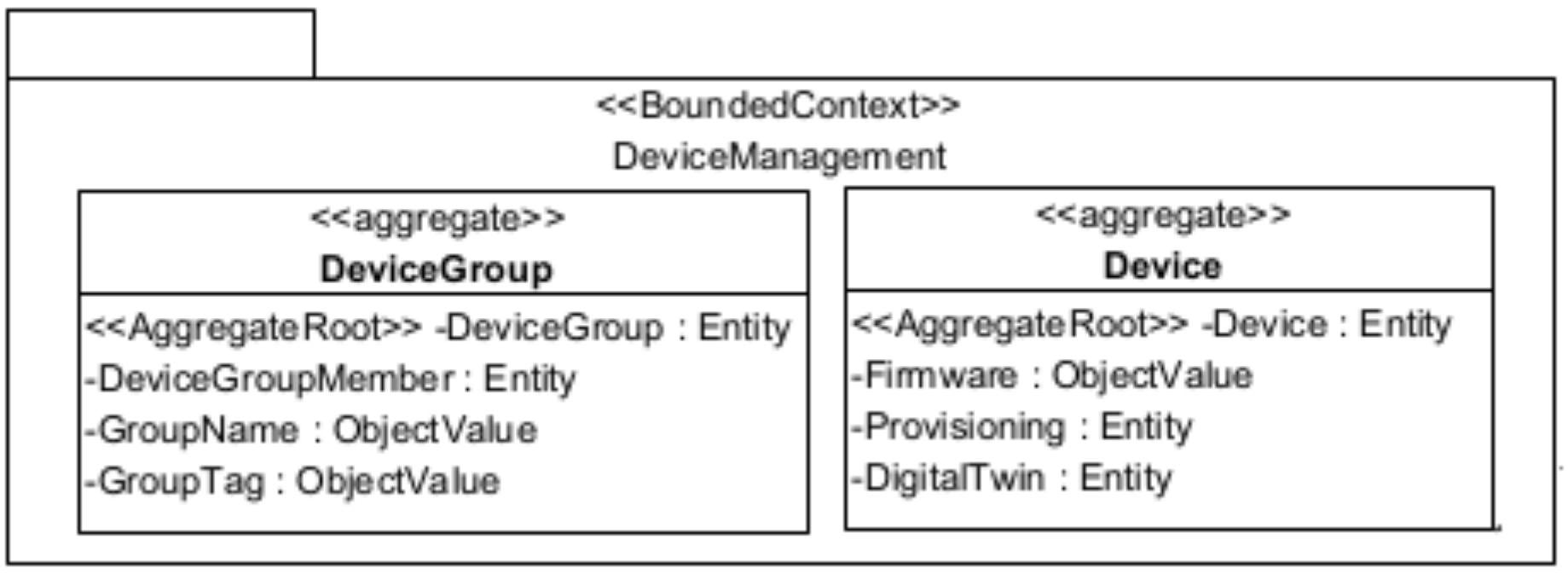

**Fig. 2.** Aggregates of IoT device management bounded context.

- **Firmware (Value Object):** An immutable structure defined by Version, Checksum, and ReleaseDate. State updates are achieved by instantiating and assigning a completely new Firmware Value Object instance rather than mutating child properties.
- **HardwareSpec (Value Object):** An immutable description characterizing physical device boundaries, encapsulating properties such as HardwareRevision, MacAddress, and RamSize.

*Device Group Aggregate*

Designed to facilitate fleet orchestration and bulk configuration actions across massive device clusters simultaneously.

- **DeviceGroup Entity (Aggregate Root):** Enforces organizational hierarchy, uniform naming invariants, and validation logic over collections of devices.
- **DeviceGroupMember Entity:** To prevent over-allocation and maintain clean lifecycle boundaries, this entity stores lightweight references (DeviceId) rather than instantiating nested Device objects inside the group fleet.
- **Group Metadata (Value Objects):** Immutable descriptors validating group identity upon mutation or creation, including GroupName, GroupDescription, and GroupTag.

*Domain Events*

Reactive, event-driven communication originating from this context is driven via explicit events:

- **DeviceRegistered**: Triggered immediately upon successful device onboarding.
- **DeviceStateChanged:** Fired when a telemetry or status attribute updates within the virtual DigitalTwin.
- **FirmwareUpdateInitiated** / **FirmwareUpdateCompleted**: Encapsulates deployment lifecycles.
- **ConnectionLost** / **ConnectionRestored**: Signals network topology fluctuations.

**Data Ingestion Bounded Context**

The Data Ingestion Bounded Context focuses exclusively on the highly scalable collection, validation, protocol conversion, and transformation of heterogeneous inbound streaming data. It decouples data-shaping mechanisms from downstream core business engines.

*Gateway Aggregate*

Acts as the architectural edge mediator bridging raw physical metrics and normalized system ingestion models.

- **Gateway Entity (Aggregate Root):** Serves as the central interface connecting physical edge hardware layers to the downstream Monitor components. It handles multi-device stream consolidation and serves as the gateway entry point.
- **DataAggregation Entity:** Processes incoming telemetry packets structurally, executing high-throughput, low-latency normalization and compression routines to format raw inputs into expected application layouts.
- **ProtocolTranslator Entity:** Functions as an Anti-Corruption Layer at the system boundary. It receives telemetry from diverse field hardware in native languages (e.g., Bluetooth, Zigbee) and normalizes it into a standardized internet format (e.g., MQTT or JSON) before forwarding payloads upward.
- **TelemetryStream Entity:** Captures and serializes raw inbound data packets emitted from physical devices prior to validation processing.

Fig. 3 depicts the Data Ingestion Bounded Context using a UML package diagram.

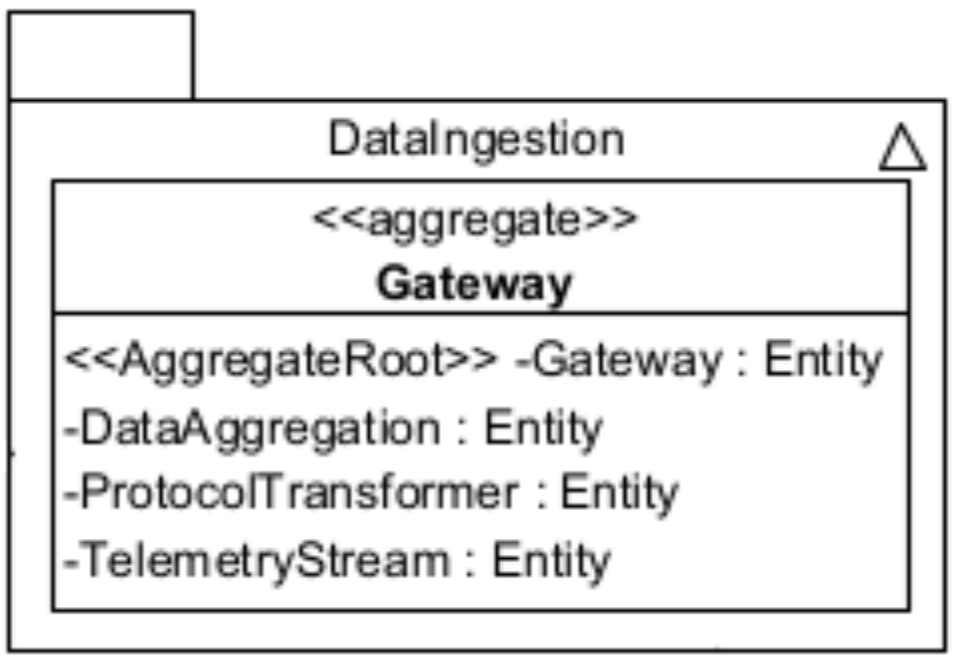

**Fig. 3.** IoT Data Ingestion Bounded Context.

**Application Business Logic Bounded Context**

This context orchestrates core application business workflows, mapping unique processes to high-level system requirements, such as ambient HVAC regulation in a Smart Home deployment or automated localized dosing in Smart Healthcare environments.

*Business Process Aggregates*

Represent highly application-specific components. Because domain invariants are strictly tied to particular localized use cases, these aggregates vary across deployments and are modeled uniquely to fit the target enterprise domain.

**Monitor Bounded Context**

Coordinates the initial observation phase of the autonomic loop, defining runtime observational policies.

- **Monitor Entity (Aggregate Root):** Establishes an asynchronous Master-Slave architecture where the Monitor acts as a supervisor observing streaming data from underlying Gateway instances. Programmatically, this is executed via a standard Observer design pattern.
- **SystemState Entity:** Evaluates the aggregate values of active system components alongside localized environmental parameters to represent the absolute runtime context, signaling adjustments when deviations from desired conditions occur.
- **MonitoringMode Entity:** Governs the sampling policy through two operational vectors based on temporal and event-driven parameters:
    - *Active Mode:* Operating proactively, the context suspends systematic metric propagation, dispatching notifications to the Analyze context strictly upon anomaly detection or state-boundary exceptions.
    - *Passive Mode:* Executes a deterministic, time-triggered paradigm, consistently generating and transmitting state vectors at predefined intervals (e.g., every 20 seconds).

Fig. 4 shows the Monitor bounded context using a UML package diagram.

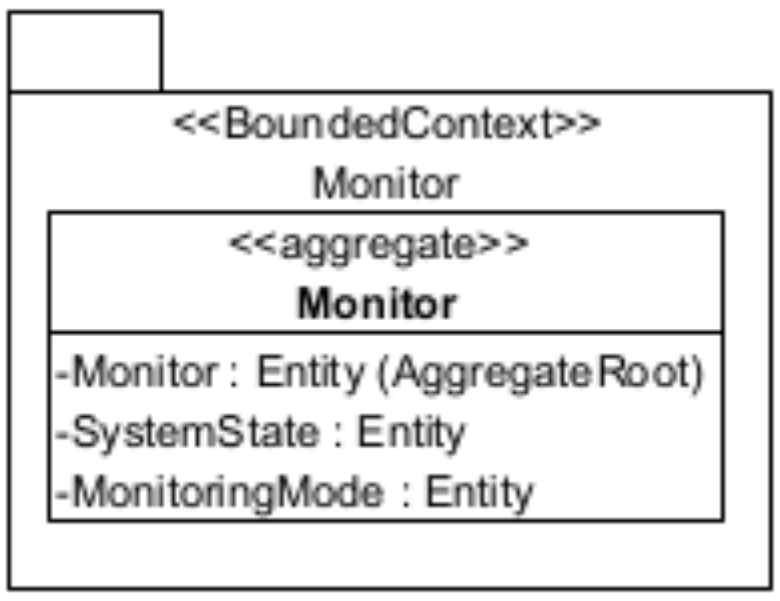

**Fig. 4.** IoT Monitor Bounded Context.

**Analyze Bounded Context**

Evaluates chronological system state metrics against established non-functional criteria and SLA specifications.

**Analyzer Entity (Aggregate Root):** The primary processing pipeline executing statistical modeling and boundary checks against streaming metrics received from the monitoring layer.

- **Symptom Entity:** Defines specific undesirable system states (e.g., highly loaded servers or network bottlenecks) through combinations of logical rules. When internal predicates are fulfilled, it flags an anomaly.
- **AdaptationRequest Entity:** An ephemeral transaction token generated when a Symptom is verified. It aggregates relevant diagnostic data and signals the planning layer asynchronously.

Fig. 5 shows the Analyze bounded context using a UML package diagram.

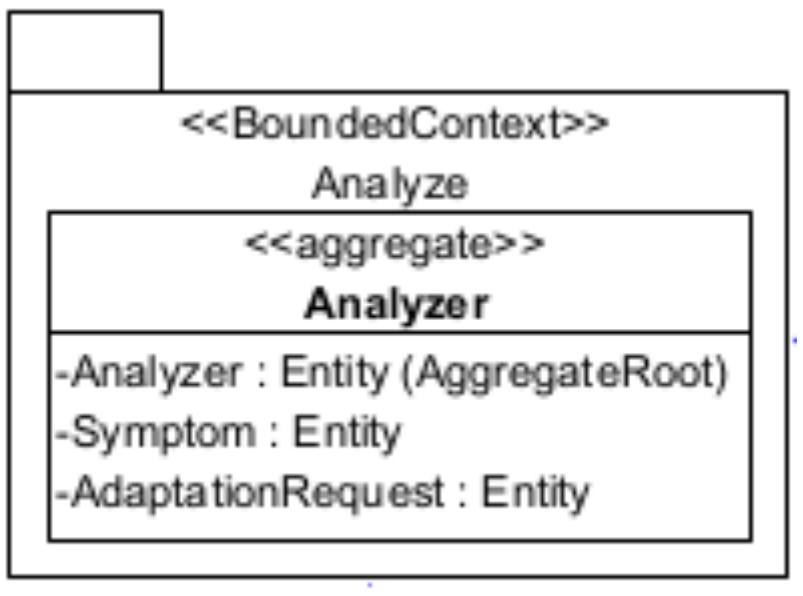

**Fig. 5.** IoT Analyze Bounded Context.

**Plan Bounded Context**
Responsible for computing runtime configuration modifications in response to flagged adaptation tokens.

- **Planner Entity (Aggregate Root):** Interacts directly with localized policy and rule evaluation engines to synthesize alternative deployment strategies.
- **ChangeStrategyPlan Entity:** An ordered sequence of explicit system mutations (e.g., auto-scaling a microservice container or shifting edge routing tables) structured to perform adaptation steps in a logical sequence.

Fig. 6 depicts the Plan bounded context using a UML package diagram.

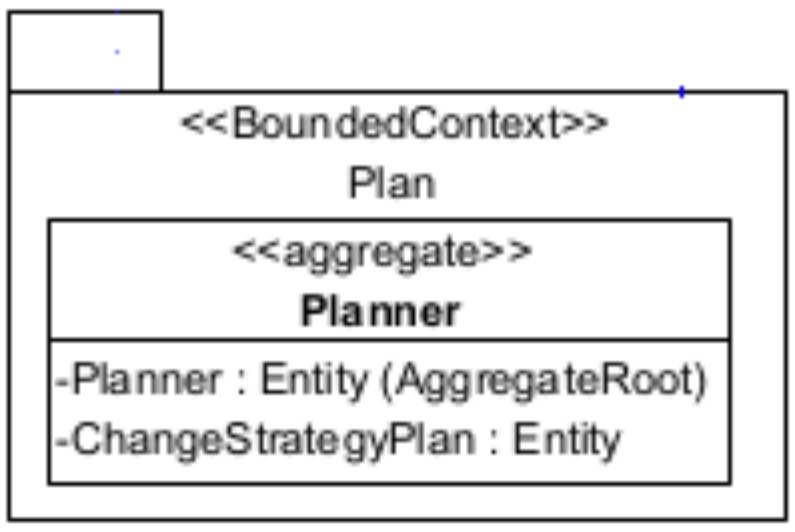

**Fig. 6.** IoT Plan Bounded Context.

**Execute Bounded Context**
Acts as the mechanical enforcement engine implementing calculated change blueprints.

- **Executor Entity (Aggregate Root):** Translates declarative change steps into low-level execution instructions, interacting with a matrix of virtual or physical effectors and actuators to apply modifications to system parameters without service disruption.

Fig. 7 depicts the Execute bounded context using a UML package diagram.

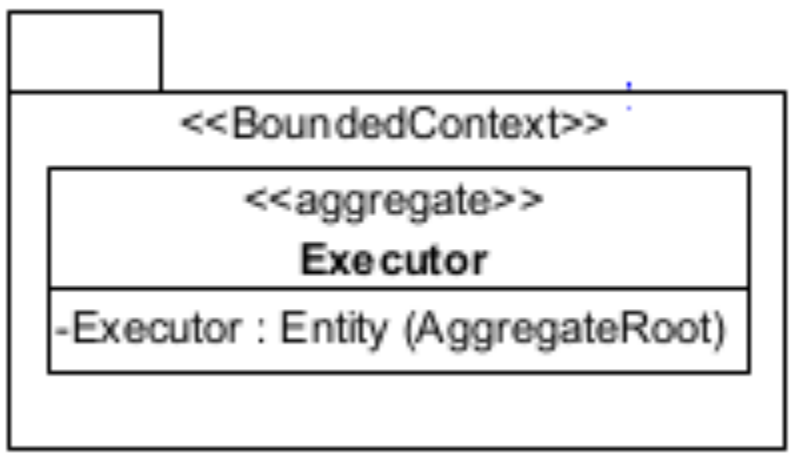

**Fig. 7.** IoT Execute Bounded Context.

**Knowledge Base Bounded Context**
Serves as the shared, distributed data substrate that provides semantic consistency across distinct stages of the autonomic control loop.

- **History Aggregate:** Manages long-term telemetry archiving, logging, and historical trend paths for trend analysis.
- **DigitalTwin Mirror Aggregate:** A read-optimized, stateful representation that maintains active operational metrics and real-time telemetry of the device fleet. Unlike the operational DigitalTwin entity in the Device Management boundary, this component does not process inbound management commands or enforce hardware-level business rules. Instead, it operates as a scalable data cache designed to quickly serve the frequent query demands of the Monitor and Analyze execution loops. By separating this analytical data path from core device management operations, the system ensures that intensive monitoring evaluations do not cause performance degradation or resource conflicts within the primary device database.
- **RuleEngine & Symptoms Aggregates:** Formalize policy rules and analytical predicates, exposing them to the Plan and Analyze boundaries as an authoritative Published Language**.**

Aggregates of the Knowledge Base bounded context are depicted in fig. 8.

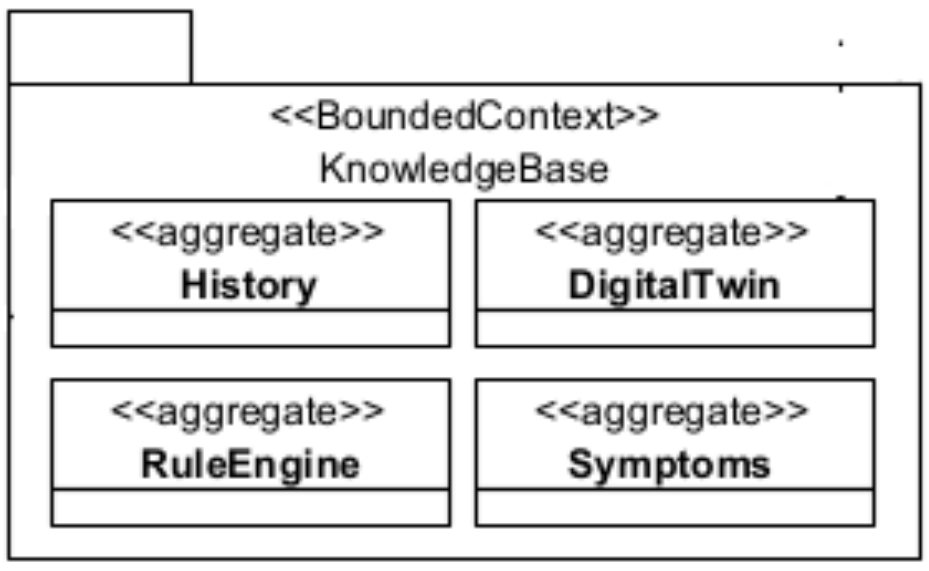

**Fig. 8.** IoT Knowledge Base bounded context.

### 4.3 Strategic IoT Context Map

To govern downstream communication channels and dependencies between isolated microservices, the system framework formalizes interactions through a structured Context Map. Instead of relying on generic data-flow connections, relationships between Bounded Contexts are categorized using standard DDD relationship typologies to ensure architectural isolation and maintain structural boundaries**.**

**Device Management to Data Ingestion Interaction**

The Device Management bounded context acts as an Upstream service provider to the Data Ingestion bounded context. It exposes its capabilities via an Open Host Service (OHS) interface. This configuration allows edge gateways within the ingestion layer to verify device registrations, validate identities, and check firmware constraints without gaining access to the internal lifecycle logic of the device registry.

**Data Ingestion to Monitor (Customer-Supplier)**

The relationship between Data Ingestion and the autonomic Monitor context follows a Customer-Supplier pattern. The Data Ingestion gateway acts as the supplier, collecting, translating, and normalizing raw physical metrics from edge networks. It then passes these standardized event packets down to the monitoring engine, which acts as the customer relying on this stream to construct the system state view.

**Monitor, Analyze, and Plan to Knowledge Base**

The Knowledge Base context operates as a Shared Kernel across autonomic MAPE-K loop components. Because the monitoring, analysis, and planning stages must access a synchronized view of system states, rules, and historical logs, they share this common data core. Any modification to core data models within the kernel requires mutual agreement across these context teams, ensuring semantic consistency across the feedback loop.

**Plan to Execute (Anti-Corruption Layer)**

The Plan context operates Upstream from the Execute context, delivering synthesized adaptation plans when system adjustments are required. To protect high-level planning logic from specific technical details of physical actuators or virtual container environments, the Execute context implements an Anti-Corruption Layer (ACL) **..** This ACL serves as a translator, mapping declarative planning goals into low-level configuration scripts or hardware commands, ensuring that infrastructure changes do not pollute core architectural logic.

**Asynchronous Synchronization via an Event-Driven Anti-Corruption Layer**

To maintain a strict separation between Core IoT Domains (Managed System) and Self-Adaptation Domains (Managing System), synchronization between the operational Device aggregate and the analytical DigitalTwin Mirror aggregate is managed via an Event-Driven Anti-Corruption Layer (ACL).

Rather than using direct database sharing or synchronous messaging—which would create tight dependencies between contexts—the synchronization process operates asynchronously:

- **Event Ingestion:** When the operational DigitalTwin entity inside the Device Management context updates its state, the aggregate root dispatches a DeviceStateChanged domain event containing relevant hardware data and telemetry.
- **Boundary Protection (The ACL):** The Knowledge Base context receives this event through a dedicated Anti-Corruption Layer (ACL) adapter at its perimeter. This ACL acts as a translator, extracting raw telemetry fields and converting them into the specific format required by the Knowledge Base ubiquitous language.
- **Semantic Isolation:** This translation prevents internal analytics, reasoning rules, and data structures of the MAPE-K loop from becoming directly dependent on registration details or specific lifecycle properties used inside the Device Management context.
- **Consistency:** Once transformed by the ACL, data updates the local DigitalTwin Mirror aggregate. This allows the managing loop to achieve eventual consistency with minimal latency while avoiding performance penalties of distributed data.

Figure 9 depicts the proposed IoT context map which illustrates the interaction between the different bounded contexts modelled in this reference architecture.

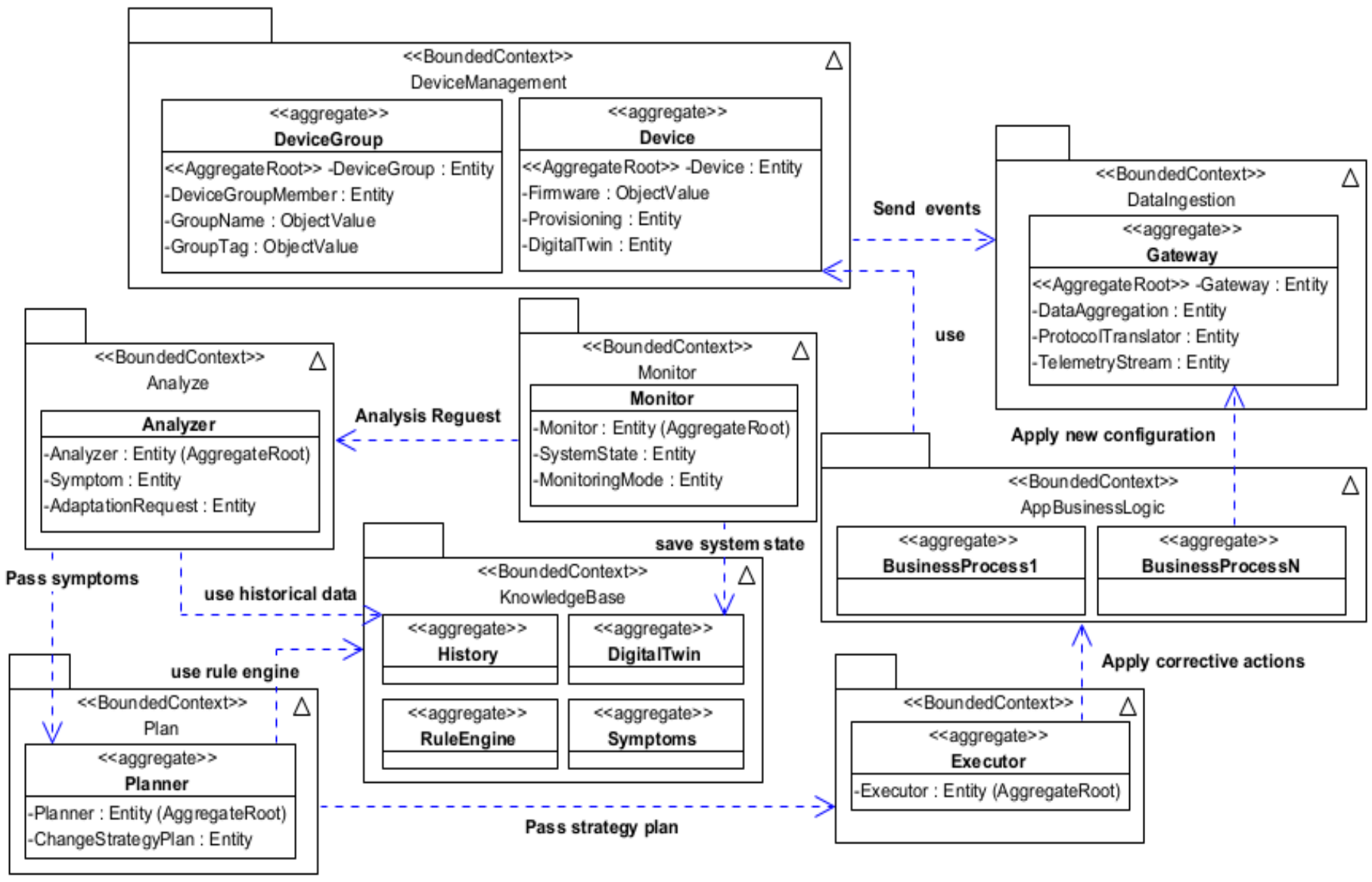

**Fig. 9.** A Context Map for MAPE-K based IoT Systems.

## 5 Conclusion and Future Work

This paper presents a novel approach utilizing Domain-Driven Design (DDD) for the engineering of self-adaptive Internet of Things (IoT) systems, successfully mitigating the "environmental monolith" anti-pattern. Foundational subdomains were identified based on the separation of concerns principle, modeling the overall environment as a self-adaptive architecture divided into a managing and a managed subsystem. Projecting these subdomains onto distinct Bounded Contexts provided a rigorous mechanism for defining microservice boundaries. This reference architecture demonstrates that adopting a structured DDD methodology brings critical structural benefits to the IoT system designer:

- **Mitigation of the Conceptual Gap:** The establishment of a localized Ubiquitous Language bridges the terminology gap between domain experts and software engineers, ensuring unambiguous stakeholder communication and aligning software implementations with target business logic.
- **Enhanced Resilience and Horizontal Scalability:** Treating the Tactical Aggregate as a first-class citizen early in the design phase establishes strict consistency boundaries. This structural isolation is critical for achieving the horizontal scalability required to process high-velocity, big data streams without system degradation.
- **Quality Attribute Realization:** Placing explicit analytical focus directly on core domain realities rather than on complex, shifting workflows enables cross-functional teams to isolate functionality, realizing crucial non-functional attributes such as modifiability, elasticity, and scalability.

While this reference architecture establishes a robust baseline for self-adaptive IoT engineering, several prospective research avenues remain open. Future work will address the following empirical and operational dimensions:

- **Detailed Microservices Evaluation:** Developing a fully realized microservices-based deployment to comprehensively evaluate the reference architecture's feasibility, highlight critical runtime design decisions, and empirically benchmark quality attributes such as ingestion latency and broker throughput.
- **Localized Context Exploration:** Conducting isolated, in-depth investigations into each proposed Bounded Context to further refine their internal tactical patterns, domain boundaries, and data-shaping mechanisms.
- **Design Pattern Cataloging:** Formulating a comprehensive catalog of software design patterns and integration templates necessary to guide engineering teams during the programmatic implementation of the proposed architecture.
- **Edge-Assisted and Intelligent Control Loops:** Offloading centralized MAPE-K execution steps into decentralized control loops running directly on distributed edge gateways, while exploring Reinforcement Learning (RL) models within the planning layers to automate context-aware adaptation policies.

# معمارية مرجعية مبنية على التصميم الموجه بالمجال لهندسة أنظمة إنترنت الأشياء

**يوسف أبوستة**

كلية تقنية المعلومات، جامعة طرابلس ،طرابلس، ليبيا
y.abusetta@uot.edu.ly

**المستخلص.** ما يزال نموذج إنترنت الأشياء يكتسب زخمًا كبيرًا في مجالات متنوعة، مدفوعًا في المقام الأول بفوائده التحولية والاقتصادية والاجتماعية. ومع ذلك، فإن هندسة نظم قوية لإنترنت الأشياء لا تزال تشكل تحديًا هائلاً؛ ويرجع ذلك إلى حد كبير إلى الديناميكية المتأصلة، والتقلب، وعدم اليقين في بيئات تشغيلها. وللتخفيف من هذا التعقيد المعماري، برزت الخدمات المصغرة (Microservices) كبديل ناجع للأطر الأحادية التقليدية الصلبة (Monolithic Frameworks) ، مما يوفر آلية قابلة للتوسع لتلبية خصائص الجودة الحرجة (Quality Attributes) ولتحقيق أقصى قدر من الفعالية لمعمارية الخدمات المصغرة، يجب أن تتوافق حدودها التقنية بدقة مع متطلبات العمل الخاصة بالمجال .(Business Requirements) ويعالج التصميم الموجه بالمجال (DDD) هذه الحاجة من خلال التفكيك الاستراتيجي للمجالات المعقدة إلى مجالات فرعية مبيّنة (Subdomains) وقابلة للإدارة العالية، والتي تطابق مباشرة الخدمات المصغرة الفردية. علاوة على ذلك، ولتقليل الاقتران المعماري (Architectural Coupling) ، وتعزيز قابلية تكامل النظام، وتحسين كفاءة استهلاك الطاقة للأجهزة، تعمل المعمارية الموجهة بالحدث (EDA) كممكّن تشغيلي حاسم. تقترح هذه الورقة معمارية مرجعية (Reference Architecture) شاملة لهندسة نظم إنترنت الأشياء، وتستفيد بشكل منهجي من مبادئ (DDD) و (EDA) لتحقيق الموائمة بين منطق العمل (Business Logic) والخدمات المصغرة المستقلة (Decoupled) والموجهة بالأحداث.